\documentstyle[12pt]{article}
\begin{document}
\title{Duality and Cosmology}
\author{B.G. Sidharth\\
B.M. Birla Science Centre, Adarshnagar, Hyderabad - 500 063, India}
\date{}
\maketitle
\begin{abstract}
In some recent theories including Quantum SuperString theory we encounter duality -
it arises due to a non commutative geometry which in effect adds an extra term
to the Heiserberg Uncertainity Principle. The result is that the micro world
and the macro universe seem to be linked. We show why this is so in the context
of a recent cosmological model and a physical picture emerges in the context
of the Feynman-Wheeler formulation of interactions.
\end{abstract}
\section{Introduction}
Nearly a century ago several Physicists including Lorentz, Poincare
and Abraham amongst others tackled unsuccessfully the problem of
the extended electron\cite{r1,r2}. An extended electron
appeared to contradict Special Relativity, while on the other hand, the limit
of a point particle lead to inexplicable infinities. Dirac finally formulated
an equation in which the physically relevant or "renormalized" mass was finite
and consisted of the bare mass and the electromagnetic mass which become
infinite in the limit of point particles, no doubt, but the infinities cancel
one another. This approach lead to non-causal effects, which were circumvented
by a formalism of Feynam and Wheeler, in which the interaction of a charge
with the rest of the universe was considered, and also not just the point
charge, but its neighbourhood had to be taken into account.\\
These infinities persisted for many decades. Infact the Heisenberg Uncertainity Principle straightaway
leads to infinities in the limit of spacetime points. It was only through
the artifice of renormalization that 't Hooft could finally circumvent this
vexing problem, in the 1970s.\\
Nevertheless it has been realized that the concept of spacetime points is only
approximate\cite{r3,r4,r5,r6,r7}. We are beginning to realize that it may be
more meaningful to speak in terms of spacetime foam, strings, branes, non
commutative geometry, fuzzy spacetime and so on\cite{r8}. This is what we
will now discuss.
\section{Duality}
We consider the well known theory of Quantum SuperStrings and also an
approach in which an electron is considered to be a Kerr-Newman Black Hole,
with the additional input of fuzzy spacetime.\\
As is well known, String Theory originated from phenomenological considerations
in the late sixties through the pioneering work of Veneziano, Nambu and
others to explain features like the s-t channel dual resonance scattering and
Regge trajectories\cite{r9}. Originally strings were conceived as one
dimensional objects with an extension of the order of the Compton wavelength,
which would fudge the point vertices of the s-t channel scattering graphs,
so that both would effectively correspond to one another (Cf.ref.\cite{r9}).\\
The above strings are really Bosonic strings. Raimond\cite{r10}, Scherk\cite{r11}
and others laid the foundation for the theory of Fermionic strings. Essentially
the relativistic Quantized String is given a rotation, when we get back the
equation for Regge trajectories,
\begin{equation}
J \leq (2\pi T)^{-1}M^2 + a_0\hbar \quad \mbox{with}\quad a_0 = +1(+2)
\mbox{for the open (closed) string}\label{e6}
\end{equation}
When $a_0 = 1$ in (\ref{e6}) we have gauge Bosons while
$a_0 = 2$ describes the gravitons. In the full theory of Quantum Super Strings, or QSS,
we are essentially dealing with extended objects rotating with the velocity of
light, rather like spinning black holes. The spatial extention is at the
Planck scale while features like extra space time dimensions which are curled
up in the Kaluza Klein sense and, as we will see, non commutative geometry
appear\cite{r12,r13}.\\
We next observe that it is well known that the
Kerr Newman of charged spinning Black Hole itself mimics the electron remarkably well including
the purely Quantum Mechanical anomalous $g=2$ factor\cite{r14}. The problem is
that there would be a naked singularity, that is the radius would become complex,
\begin{equation}
r_+ = \frac{GM}{c^2} + \imath b, b \equiv \left(\frac{G^2Q^2}{c^8} + a^2 -
\frac{G^2M^2}{c^4}\right)^{1/2}\label{e7}
\end{equation}
where $a$ is the angular momentum per unit mass.\\
This problem has been studied in detail by the author in recent years\cite{r15,r16}.
Indeed it is quite remarkable that the position coordinate of an electron in
the Dirac theory is non Hermitian and mimics equation (\ref{e7}), being
given by
\begin{equation}
x = (c^2 p_1 H^{-1}t+a_1) + \frac{\imath}{2} c\hbar (\alpha_1 - cp_1H^{-1})H^{-1},\label{e8}
\end{equation}
where the imaginary parts of (\ref{e7}) and (\ref{e8}) are both of the order
of the Compton wavelength.\\
The key to understanding the unacceptable imaginary part was given by Dirac
himself\cite{r17}, in terms of zitterbewegung. The point is that according to
the Heisenberg Uncertainity Principle, space time points themselves are not
meaningful- only space time intervals have meaning, and we are really speaking of
averages over such intervals, which are atleast of the order of the Compton
scale. Once this is kept in mind, the imaginary term disappears on averaging
over the Compton scale.\\
In this formulation, the mass and charge of the electron arises due to
zitterbewegung effects at the Compton scale\cite{r15,r16}. These masses and
charges are renormalized in the sense of the Dirac mass in the classical
theory, alluded to in section 1.\\
Indeed, from a classical point of view also, in the extreme relativistic
case, as is well known there is an extension of the order of the Compton
wavelength, within which we encounter meaningless negative energies\cite{r18}.
With this proviso, it has been shown that we could think of an electron as a
spinning Kerr Newman Black Hole. This has received independent support from
the work of Nottale\cite{r19}.\\
We are thus lead to the picture where there is a cut off in space time
intervals.\\
In the above two scenarios, the cut off is at the Compton scale $(l,\tau)$ the
Planck scale being a special case for the Planck mass.
Such discrete space time models compatible with Special Relativity have been
studied for a long time by Snyder and several other scholars\cite{r20,r21,r22}.
In this case it is well known that we have the following non commutative
geometry
$$[x, y] = (\imath a^2 / \hbar)L_{z,}  [t, x] = (\imath a^2 / \hbar c)M_{x,}$$
\begin{equation}
[y, z] = (\imath a^2 / \hbar) L_{x,} [t, y] = (\imath a^2 / \hbar c)M_{y,}\label{e9}
\end{equation}
$$[z, x] = (\imath a^2 / \hbar) L_{y,} [t, z] = (\imath a^2 / \hbar c)M_{z,}$$
where $a$ is the minimum natural unit and $L_x, M_x$ etc. have their usual significance.\\
Moreover in this case there is also a correction to the usual Quantum Mechanical
commutation relations, which are now given by
$$[x, p_x] = \imath \hbar [1+(a/\hbar)^2 p^2_x];$$
$$[t, p_t] = \imath \hbar [1-(a/ \hbar c)^2 p^2_t];$$
\begin{equation}
[x, p_y] = [y, p_x] = \imath \hbar (a/ \hbar)^2 p_xp_y ;\label{e10}
\end{equation}
$$[x, p_t] = c^2[p_{x,} t] = \imath \hbar (a/ \hbar)^2 p_xp_t ;\mbox{etc}.$$
where $p^\mu$ denotes the four momentum.\\
In the Kerr Newman model for the electron alluded to above (or generally
for a spinning sphere of spin $\sim \hbar$ and of radius $l$), $L_x$ etc. reduce
to the spin $\frac{\hbar}{2}$ of a Fermion and the commutation relations
(\ref{e9}) and (\ref{e10}) reduce to
\begin{equation}
[x,y] \approx 0(l^2),[x,p_x] = \imath \hbar [1 + \beta l^2], [t,E] = \imath \hbar [1+\tau^2]\label{e11}
\end{equation}
where $\beta = (p_x/\hbar)^2$ and similar equations.\\
Interestingly the non commutative geometry given in (\ref{e11}) can be shown
to lead to the representation of Dirac matrices and the Dirac equation\cite{r23}.
From here we can get the Klein Gordon equation, as is well known\cite{r24,r25},
or alternatively we deduce the massless string equation.\\
This is also the case with superstrings where Dirac spinors are
introduced, as indicated above. Infact in QSS also we have equations
mathematically identical to the relations (\ref{e11}) containing
momenta (Cf.ref.\cite{r13}). This, which implies (\ref{e9}), can
now be seen to be the origin of non-commutativity.\\
The non commutative geometry and fuzzyness is contained in (\ref{e11}).
Infact fuzzy spaces have been investigated in detail by Madore and others\cite{r26,r27},
and we are lead back to the equation (\ref{e11}). The fuzzyness which is
closely tied up with the non commutative feature is symptomatic of the breakdown
of the concept of the spacetime points and point particles at small scales
or high energies. As has been noted by Snyder, Witten, and several other
scholars, the divergences encountered in Quantum Field Theory are symptomatic
of precisely such an extrapolation to spacetime points and which necessitates
devices like renormalization. As Witten points out\cite{r28}, "in developing
relativity, Einstein assumed that the space time coordinates were Bosonic;
Fermions had not yet been discovered!... The structure of space time is
enriched by Fermionic as well as Bosonic coordinates."\\
A related concept, which one encounters also in String Theory is Duality. Infact
the relation (\ref{e11}) leads to,
\begin{equation}
\Delta x \sim \frac{\hbar}{\Delta p} + \alpha' \frac{\Delta p}{\hbar}\label{e18}
\end{equation}
where $\alpha' = l^2$, which in Quantum SuperStrings Theory $\sim 10^{-66}$. Witten
has wondered about the basis of (\ref{e18}), but as we have seen, it is a
consequence of (\ref{e11}).\\
This is an expression of the duality relation,
$$R \to \alpha'/R$$
This is symptomatic of the fact that we cannot go down to arbitrarily small
spacetime intervals, below the Planck scale.\\
There is an interesting meaning to the duality relation arising from (\ref{e18}) in
the context of the Kerr-Newman Black Hole formulation.
While it appears that the ultra small is a gateway to the macro cosmos, we
could look at it in the following manner. The first term of the relation
(\ref{e18}) which is the usual Heisenberg Uncertainity relation is supplemented
by the second term which refers to the macro cosmos.\\
Let us consider the second term in (\ref{e18}). We write $\Delta p = \Delta Nmc$,
where $\Delta N$ is the Uncertainity in the number of particles, $N$, in the
universe. Also $\Delta x = R$, the radius of the universe where
\begin{equation}
R \sim \sqrt{N}l,\label{ex}
\end{equation}
the famous Eddington relationship. It should be stressed that the otherwise
emperical Eddington formula, arises quite naturally in a Brownian characterisation
of the universe as has been pointed out earlier (Cf. for example ref.\cite{r5}). Put
simply (\ref{ex}) in the Random Walk equation\\
We now get, using (\ref{e7}),
$$\Delta N = \sqrt{N}$$
Substituting this in the time analogue of the second term of (\ref{e18}), we
immediately get, $T$ being the age of the universe,
\begin{equation}
T = \sqrt{N} \tau\label{ey}
\end{equation}
In the above analysis, including the Eddington formula (\ref{ex}), $l$ and $\tau$ are the
Compton wavelength and Compton time of a typical elementary particle, namely
the pion. The equation for the age of the universe is also correctly given
above. Infact in the closely related model of fluctuational cosmology (Cf. for
example ref.\cite{r29}) all of the Dirac large number coincidences as also
the mysterious Weinberg formula relating the mass of the pion to the Hubble
constant, follow as a consequence, and are not emperical. In this formulation,
in a nutshell, $\sqrt{N}$ particles are fluctuationally created within the
time $\tau$, so that,
\begin{equation}
\frac{dN}{dt} = \frac{\sqrt{N}}{\tau}\label{ez}
\end{equation}
which leads to (\ref{ey}) (and (\ref{ex})).\\
Next use of the well known formula, $(M = Nm, M$ being the mass of the universe,
and $m$ the pion mass)
$$R \approx GM/c^2,$$
gives on differentiation and use of (\ref{ez}) the Hubble law, with
\begin{equation}
H = \frac{c}{l} \frac{1}{\sqrt{N}} \approx \frac{Gm^3c}{\hbar}
\mbox{or} \quad m = \left(\frac{\hbar H}{Gc}\right)^{1/3}\label{ea}
\end{equation}
(\ref{ea}) gives the supposedly mysterious and empirical Weinberg formula connecting the pion mass to
the Hubble constant.\\
Using (\ref{ea}) we can deduce that there can be a cosmological constant
$\Lambda$ such that,
$$\Lambda \leq 0(H^2)$$
Recent observations confirm this ever expanding and possibly accelerating feature of the
universe\cite{pearl}.
All these relations
relating large scale parameters to microphysical constants were shown to be
symptomatic of what has been called, stochastic holism (Cf. also ref.\cite{r30}),
that is a micro-macro connection with a Brownian or stochastic underpinning.
Duality, or equivalently, relation (\ref{e18}) is really an expression of
this micro-macro link.\\
\section{The Dirac and Feynman - Wheeler Formulations}
To appreciate this concept of holism in a more physical sense, we return
to the classical description of the electron alluded to right at the beginning.
We will discuss very briefly the contributions of Dirac, Feynman and Wheeler. This was built upon
the earlier work of Lorentz, Abraham, Fokker and others. Our starting point
is the so called Lorentz-Dirac equation\cite{r2}:
\begin{equation}
ma^\mu = F^\mu_{in} + F^\mu _{ext} + \Gamma^\mu\label{e101}
\end {equation}
where
$$F^\mu_{in} = \frac{e}{c} F^{\mu v}_{in} v_v$$
and $\Gamma^\mu$ is the Abraham radiation reaction four vector related to
the self force and, given by
\begin{equation}
\Gamma^\mu = \frac{2}{3} \frac{e^2}{c^3} (\dot a^\mu - \frac{1}{c^2} a^\lambda
a_\lambda v^\mu)\label{e102}
\end{equation}
Equation (\ref{e101}) is the relativistic generalisation for a point electron of
an earlier equation proposed by Lorentz, while equation (\ref{e102}) is the
relativisitic generalisation of the original radiation reaction term due to
energy loss by radiation. It must be mentioned that the mass $m$ in equation
(\ref{e101}) consists of a neutral mass and the original electromagnetic mass
of Lorentz, which latter does tend to infinity as the electron shrinks to a
point, but, this is absorbed into the neutral mass. Thus we have the forerunner
of renormalisation in quantum theory.\\
There are three unsatisfactory features of the Lorentz-Dirac
equation (\ref{e101}).\\
Firstly the third derivative of the position coordinate in (\ref{e101})
through $\Gamma^\mu$ gives
a whole family of solutions. Except one, the rest of the solutions are run away -
that is the velocity of the electron increases with time to the velocity of
light, even in the absence of any forces. This energy can be thought to come
from the infinite self energy we get when the size of the electron shrinks
to zero. If we assume adhoc an
asymptotically vanishing acceleration then we get a physically meaningful
solution, though this leads to a second difficulty,
that of violation of causality of even the physically
meaningful solutions. Let us see this briefly.\\
For this, we notice that equation (\ref{e101}) can be written in the form\cite{r2},
\begin{equation}
ma^\mu (\tau) = \int^\infty_0 K^\mu (\tau + \alpha \tau_0)e^{-\alpha}
d\alpha\label{e103}
\end{equation}
where
$$K^\mu (\tau) = F^\mu_{in} + F^\mu_{ext} - \frac{1}{c^2}Rv^\mu,$$
\begin{equation}
\tau_0 \equiv \frac{2}{3} \frac{e^2}{mc^3}\label{e104}
\end{equation}
and
$$\alpha = \frac{\tau' - \tau}{\tau_0},$$
where $\tau$ denotes the time and $R$ is the total radiation rate.\\
It can be seen that equation (\ref{e103}) differs from the usual equation of
Newtonian Mechanics, in that it is non local in time. That is, the
acceleration $a^\mu (\tau)$ depends on the force not only at time $\tau$, but
at subsequent times also. Let us now try to characterise this non locality
in time. We observe that $\tau_0$ given by equation (\ref{e104}) is the
Compton time $\sim 10^{-23}secs.$ So equation (\ref{e103}) can be approximated
by
\begin{equation}
ma^\mu (\tau) = K^\mu (\tau + \xi \tau_0 ) \approx K^\mu (\tau)\label{e105}
\end{equation}
Thus as can be seen from (\ref{e105}), the Lorentz-Dirac equation differs from
the usual local theory by a term of the order of
\begin{equation}
\frac{2}{3} \frac{e^2}{c^3} \dot a^\mu\label{e106}
\end{equation}
the so called Schott term.\\
So, the non locality in time is within intervals $\sim \tau_0$, the Compton time exactly
what we encountered in section 2.\\
It must also be reiterated that the Lorentz-Dirac equation must be supplemented
by the asymptotic condition of vanishing acceleration in order to be
meaningful. That is, we have to invoke not just the point electron, but also
distant regions into the future as boundary conditions.\\
Finally it must be borne in mind that the four vector $\Gamma^\mu$ given in
(\ref{e102}) can also be written as
\begin{equation}
\Gamma^\mu \equiv \frac{e}{2c} (F^{\mu v}_{ret} - F^{\mu v}_{adv}) v_v\label{e107}
\end{equation}
In (\ref{e107}) we can see the presence of the advanced or acausal field which has been
considered unsatisfactory.
Infact this term, as is well known directly leads to
the Schott term (\ref{e106}). Let us examine this non local feature.
As is known, considering the time component of the Schott term (\ref{e106}) we
get (cf.ref.\cite{r2})
$$-\frac{dE}{dt} \approx R \approx \frac{2}{3} \frac{e^2c}{r^2}
(\frac{E}{mc^2})^4,$$
where $E$ is the energy of the particle.\\
whence intergrating over the period of non locality $\sim \tau_0$ we can
immediately deduce that $r$, the dimension of spatial non locality is
given by
$$r \sim c \tau_0,$$
that is of the order of the Compton wavelength. This
follows in any case in a relativistic theory, given the above Compton time. This
term represents the effects within the neighbourhood of the charge.\\
What we have done is that we have quantified the space-time interval of non
locality - it is of the order of the Compton wavelength and time. This
contains the renormalization effect and gives the correct physical
mass.\\
We now come to the Feynman-Wheeler action at a distance theory\cite{r32,r33}.
They showed that the apparent acausality of the theory would disappear if the
interaction of a charge with all other charges in the universe, such that the
remaining charges would absorb all local electromagnetic influences was
considered. The rationale behind this was that in an action at a distance
context, the motion of a charge would instantaneously affect other charges,
whose motion in turn would instantaneously affect the original charge. Thus
considering a small interval in the neighbourhood of the point charge, they
deduced,
\begin{equation}
F^\mu_{ret} = \frac{1}{2} \{ F^\mu_{ret} + F^\mu_{adv} \} + \frac{1}{2}
\{ F^\mu_{ret} - F^\mu_{adv} \}\label{e108}
\end{equation}
The left side of (\ref{e108}) is the usual causal field, while the right side has
two terms. The first of these is the time symmetric field while the second
can easily be identified with the Dirac field above and represents the sum
of the responses of the remaining charges calculated in the vicinity of the
said charge.\\
From this point of view, the self force or in the earlier Kerr-Newman formulation,
effects within the Compton scale, turns out to be the
combined reaction of the rest of the charges, or in the earlier duality and
cosmological considerations, the holistic effect.
\section{Duality and Scale}
In a previous communication\cite{r34} it was shown that we could consider a
scaled Planck constant
$$h_1 \approx N^{3/2} \hbar$$
such that we would have
$$R = \frac{h_1}{Mc}$$
It is interesting to note that these relations are essentially the same as
the second or extra term in (\ref{e18}) viz.,
$$l^2 \frac{\Delta p}{\hbar} \sim \Delta x$$
with $\Delta p = \sqrt{N} m$ and $\Delta x = R$ as before. This can be
easily verified.\\
In other words the two terms of the modified Heisenberg Uncertainity relation
(\ref{e18}) represents two scales. The first term represents the micro scale
with the Planck constant, while the second term represents the macro scale with
the scaled Planck constant $h_1$, both being linked, as noted earlier.

\end{document}